# Synthesis of Oxidation-Resistant Electrochemical-Active Copper Nanowires Using Phenylenediamine Isomers


Tan Zhang,[†] Farhad Daneshvar,[†] Cong Liu,[†] Shaoyang Wang,[‡] Mingzhen Zhao,[†] Fangqing Xia,[†] and Hung-Jue Sue[*,†]

[†] Polymer Technology Center, Department of Materials Science and Engineering, Texas A&M University, College Station, TX 77843, USA

[‡] Artie McFerrin Department of Chemical Engineering, Texas A&M University, College Station, TX 77843, USA

*Corresponding Author (hjsue@tamu.edu)




**Abstract**

Phenylenediamine (PDA) was chosen as a coordinating, reducing, and capping agent to effectively direct growth and protect against oxidation of Cu nanowires (Cu NWs) in an aqueous solution. PDA was found to reduce Cu (II) to Cu (I) at room temperature, and stabilize the resulting Cu (I) by forming a coordination complex. The presence of a stable Cu (I) complex is the key step in the synthesis of Cu NWs under mild conditions. Different PDA isomers lead to different growth paths of forming Cu NWs. Both *p*PDA and *m*PDA-synthesized Cu NWs were covered with a thin layer of polyphenylenediamine and show excellent anti-oxidation properties, even in the presence of water. The usefulness of the present and electrochemical active Cu NWs for a variety of nanotechnology applications is discussed.

## Introduction

One-dimensional nanocrystals of transition metals, such as platinum (Pt), silver (Ag) and copper (Cu) nanowires (NWs), are important conducting materials for the fabrication of electronic nano-devices. Compared with bulk metal, metal NWs exhibit unique electronic and optical properties due to their high aspect ratios and nanoscale diameters.[1-2] Among these metallic NWs, Cu NWs are of particular interest not only because of their relatively low cost but also their excellent electrical and electrochemical properties. For example, individual Cu NW has been shown to possess an ampacity nearly 40 times higher than that of bulk Cu.[3] The networks of Cu NWs showed novel electrochemical properties to use as a battery electrode and catalyst.[4-5] Cu NWs are therefore being considered as a preferred candidate material of choice for "multifunctional building block material" in nano-device fabrication.

Corrosion and oxidation is an inevitable process that occurs in metallic systems which impose significant costs to metals, from large scale industries like gas refineries to small systems like recording heads. [6-8] Recent rapid development in battery, sensor and catalytic systems requires an electrochemical material that can be stable in an aqueous environment for a prolonged lifetime. [4, 9-10] Cu NWs exhibit poor oxidation resistance at ambient conditions, especially in the presence of water or humid air, reducing their properties and lifetime if not well protected.[1, 11] Several methods have been proposed to prevent the oxidation of Cu NWs, such as deposition of a layer of protective metallic or metal oxide nanoparticles,[12-13] or wrapping them with graphene oxide.[14] Due to the rapid oxidation of Cu NWs, typically within hours,[11] these post-synthesis methods cannot effectively protect freshly prepared Cu NWs, and are not suitable for large-scale production. Forming a protective layer on Cu NWs *in-situ* during synthesis has to



be considered as an effective method to protect freshly prepared Cu NWs, which is also suggested by others.[11]

Cu NWs synthesized by the hydrothermal method are usually covered with alkyl amines. During Cu NW synthesis, these alkyl amines can coordinate with Cu ions, thereby not only promoting chemical reduction of Cu ions but also serve as a capping agent to direct one-dimensional growth of Cu nanocrystals.[15] These alkyl amines remain on the Cu NW surface after synthesis. It has been reported that long-chain alkyl amines, such as hexadecylamine (HDA), form a densely packed monolayer on Cu NW surfaces.[16] These adsorbed hydrophobic alkyl chains can repel oxygen and water molecules from attacking Cu, which slows down the oxidation of Cu NWs to some extent.[16-17] However, these long-chain alkyl amines could also inhibit the charge transfer from the solution to the Cu NW surfaces.[18] Short-chain alkyl diamines, such as ethylenediamine (EDA), have also been used to synthesize and protect Cu NWs.[11, 19-20] The Cu NWs synthesized from EDA were found to be oxidation-resistant in polar organic solvents, but oxidized easily in nonpolar organic solvents and water.[11] One solution to prepare oxidation-resistant electrochemical active Cu NWs is to have a thin layer of conducting polymer coated on Cu NWs *in-situ*.[21] However, it is difficult to find a molecule that can not only polymerize to a conducting polymer but also direct one dimensional growth of Cu nanocrystals. Up to now, the only successful attempt is using pyrrole with the assistance of a strong reducing agent hydrazine, but the stability for the resulting polypyrrole-coated Cu NWs was not tested. [21]

Herein, we report using phenylenediamine (PDA) isomers to prepare Cu NW core-shell structures. Similar to alkyl amines, the amine group(s) on PDA is able to coordinate with Cu (II) to serve as a coordinating and a capping agent during the Cu NW synthesis. However, PDA alone also can be used as a reducing agent.[22] A redox reaction has been reported to occur between PDA



and Ag (I), where Ag (I) is reduced to Ag atom, and PDA itself is oxidized and polymerized to poly(phenylenediamine) (polyPDA).[22] PolyPDA is a conducting polymer, which has been used for pseudo capacitor, dye or metal ion absorption, catalysis and biosensors.[22-25] Similar to polyaniline, polyPDA is also an efficient anticorrosion material to protect metals.[25] PolyPDA can be an effective shell to protect Cu NWs without reducing their electrochemical properties. All three isomers of PDA (*para*-, *meta*- and *ortho*-) have different oxidation potentials (*o*PDA > *p*PDA > *m*PDA),[26] Therefore, it is of interest to explore the role of PDA isomers in Cu NW synthesis for multifunctional applications.

## Experimental Section

### Materials

Copper chloride (CuCl$_2$, from Alfa Aesar), glucose (from TCI), *para*-phenylenediamine (*p*PDA, from Sigma), *meta*-phenylenediamine (*m*PDA, from Acros), *ortho*-phenylenediamine (*o*PDA, from Merck) and hexadecylamine (HDA, from Merck) were used as received.

### Cu nanowire synthesis

In the precursor solution, 0.75 mmol CuCl$_2$ and 3.0 mmol PDA with or without 0.75 mmol glucose were dissolved in 60 mL deionized water. The precursors were first stirred for 3 h at room temperature, and then kept at 110 °C for 24 h without stirring. The products were washed with deionized water and ethanol to remove unreacted reagents and polymeric byproducts. A detailed synthetic procedures for Cu NWs using HDA has been reported elsewhere.[3]

### Characterization and methods



The morphologies and the properties of the as-synthesized samples were characterized using an ultraviolet-visible near-infrared spectrophotometer (Shimadzu UV-3600), a X-ray photoelectron spectroscopy (XPS, Omicron's DAR 40), a transmission electron microscopy (JEOL JEM-2010 TEM), a scanning electron microscopy (Tescan FERA-3 focused ion beam SEM) and an X-ray powder diffractometer (Bruker D8 focus Bragg Brentano). The electrochemical properties were tested using a Solartron cyclic voltammeter by placing a Cu NWs coated ITO slide as working electrode in 0.1 M KOH solution with a scan rate of 20 mV/min. The simulation was conducted using a universal force field (UFF) in Avogadro.

## Results and Discussion

The aqueous precursor solutions containing $CuCl_2$ and PDA were stirred at room temperature for 3 h to obtain uniform suspensions. The UV-Vis spectra for the precursor solutions having different stirring time of duration were recorded and plotted in Figure 1. The absorption band for PDA isomers are located below 300 nm[27] and that for $CuCl_2$ is above 800 nm.[15] Due to a lone electron pair on the nitrogen atom, PDA can coordinate with electrophilic Cu (II) as illustrated in Figure 2. The presence of a broad absorption band centered at 560 nm suggests the formation of Cu (II)-amine occurs on mixing,[28] which is consistent with the observations for the Cu (II) coordinated with alkyl amines[15, 29]



Zhang and Sue                    Cu-PDA Hybrids

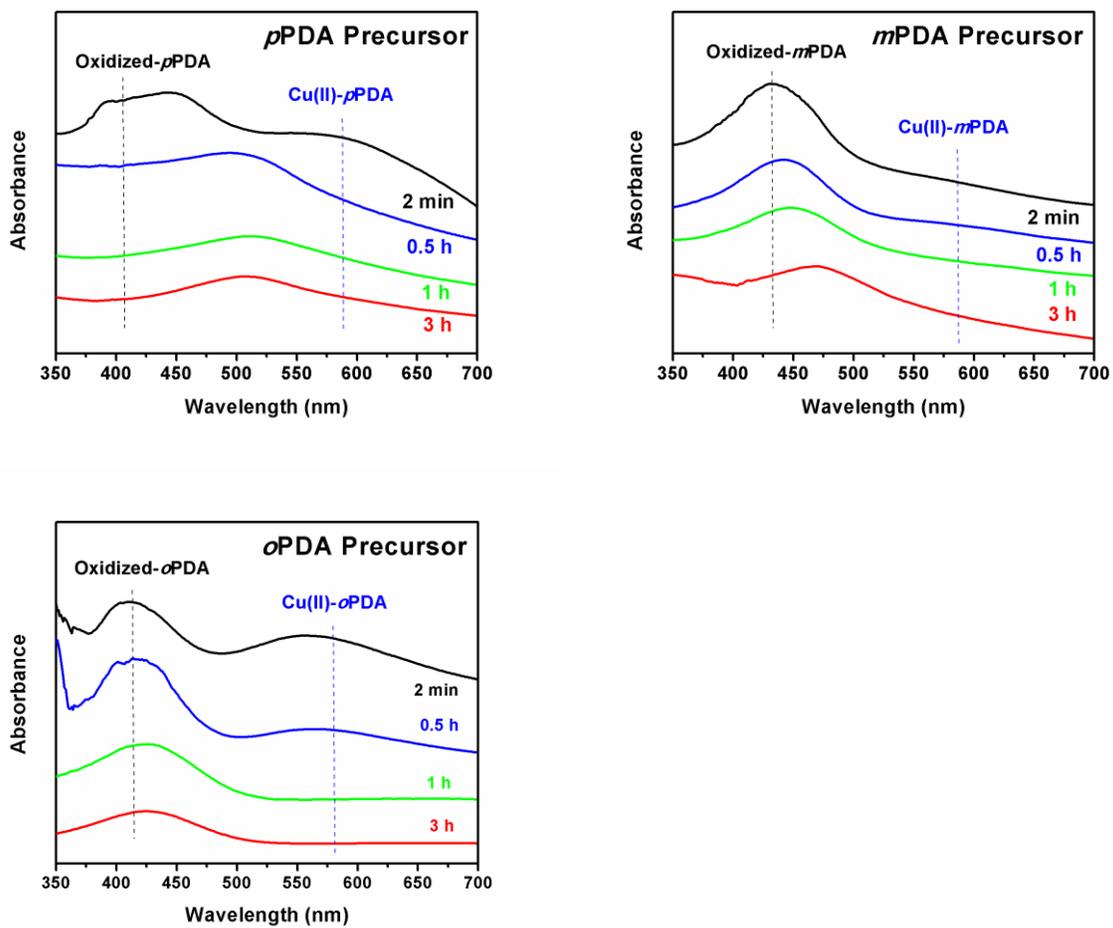

**Figure 1.** UV-Vis spectra of the precursor solutions containing *p*PDA, *m*PDA and *o*PDA.



Zhang and Sue                    Cu-PDA Hybrids

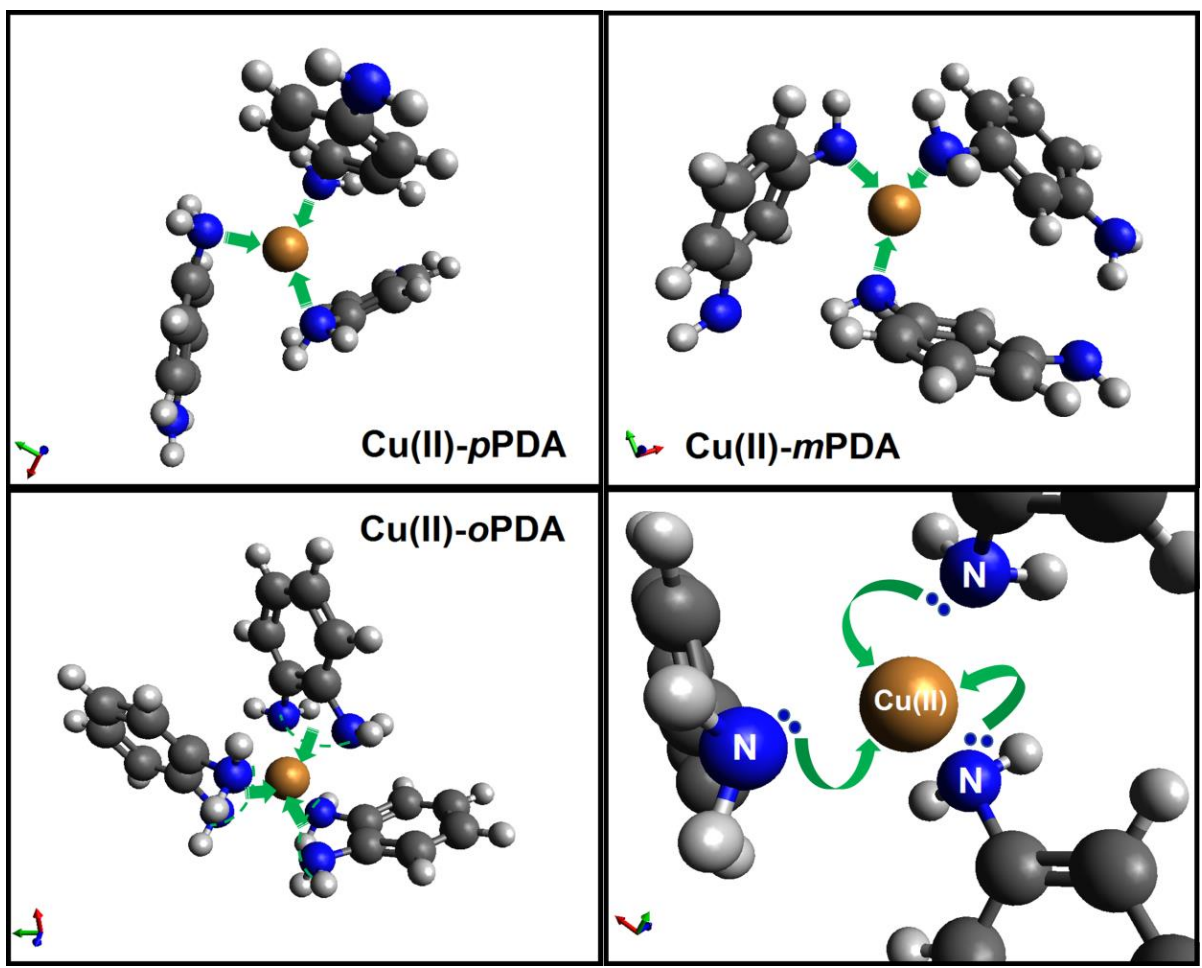

**Figure 2.** The coordination structures for the complex of Cu (II) and PDA isomers. The ball in brown, blue, dark gray and light gray color represents Cu (II), nitrogen, carbon and hydrogen atoms, respectively.

Cu (II)-PDA coordination complex is known to be unstable[30] because the electron donation from the coordinated PDA leads to a ligand-to-metal charge transfer (LMCT) where PDA is oxidized and Cu (II) is reduced. The redox reaction between Cu (II) and PDA occurs spontaneously in an aqueous environment. As seen in Figure 1, the absorption band for Cu (II)-PDA complexes at 560 nm disappeared rapidly with increased stirring time, or was not observed in the precursor solutions in the case of *m*PDA. In addition, an absorption band between 400 and 450 nm was observed within 2 min of stirring. This band is attributed to the π-π* transition from



the phenazine unit conjugated to the bridging nitrogen,[26] a typical structural characteristic for the oxidation form of PDA, suggesting that PDA was immediately oxidized by Cu (II). Oxidized PDA polymerizes,[24, 26] evidenced by the polymer formation in the precursor solution (Figure 3b). The short lifetime for Cu (II)-PDA complexes makes their structures difficult to analyze experimentally.  Figure 2 shows a 1:3 coordination ratio for Cu (II)-PDA complex based on a simulation using universal force field. The proposed structure from simulation shows that Cu (II) serves as a core, and is surrounded by three PDA molecules through the coordination bonding with their amine groups. The simulated 1:3 ratio is in line with the others' work where Cu (II) can coordinate with up to three diamines.[30]

With increased stirring time, the absorption band for the oxidized PDA is shifted towards higher wavelengths, and the shift range varies with the types of PDA isomers. The redshift for PDA absorption band suggests a higher degree of oxidation and oligomerization.[25] Based on the oxidation potential,[26] the ease of oxidation should follow the order of  $o$PDA > $p$PDA > $m$PDA. However, from Figure 2, the redshift range for the wavelength is found to be $p$PDA > $m$PDA > $o$PDA, which does not match their oxidation potential order. It is possible that some other factor also affect the redox reaction between Cu (II) and PDA other than thermodynamics solely. One possibility is the differences in their coordination structures. Due to the steric effect of aromatic ring, $m$PDA and $o$PDA can only coordinate with one Cu (II) ion. For $p$PDA, in addition to form the coordination structures as shown in Figure 2, the $para$ positioned amine group is also active, and can coordinate with the second Cu (II) ion.[30-31] This bridging structure can form stable colloids between oxidized $p$PDA and Cu ions,[31] which may further enhance the extent of redox reaction in between. The coordination structure likely also affects the ease of the redox reaction between Cu (II) and PDA.



Although PDA has been oxidized in the precursor solution, Cu (0) was not observed after 3 h of stirring at room temperature (Figure S1), which is consistent with others' work.[31-33] The suspended solids in the precursor solutions were collected and analyzed using XPS. As seen in Figure 3a, the suspended solids in the precursor containing *m*PDA or *o*PDA show two groups of peaks at (932.0 w/ 952.0 eV) and (934.7 w/ 954.3 eV). The former is attributed to Cu (I) and the latter originates from Cu (II).[34-35] By comparing the area under the each peak (the details can be found in *Supporting Information*), only half of Cu (II) was reduced to Cu (I) by *m*PDA and *o*PDA at room temperature, respectively. For the precursor containing *p*PDA, more than 85% of Cu (II) was reduced to Cu (I), which is agreed with the UV-Vis spectra in which the redox reaction in *p*PDA precursor is much more intense compared with the other two isomers.

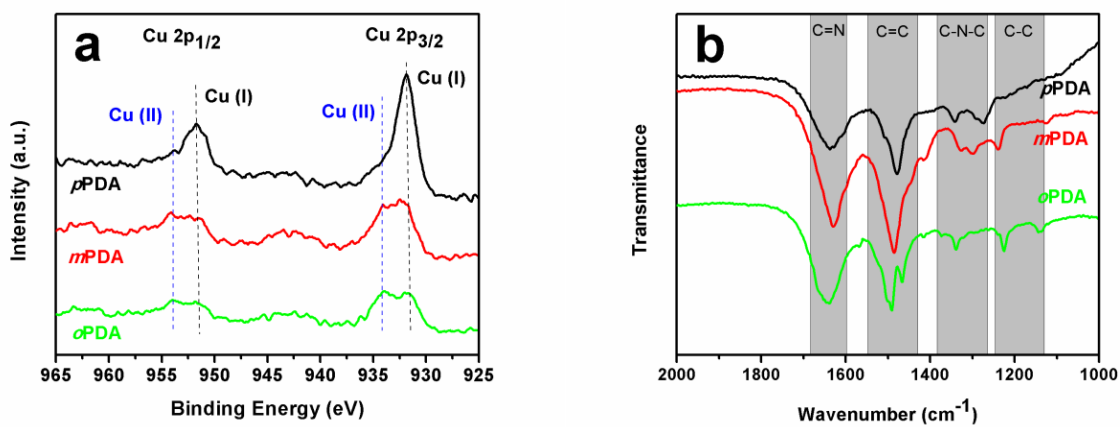

**Figure 3.** (a) XPS spectra and (b) FTIR spectra of the suspended solids in the precursor solution containing *p*PDA, *m*PDA and *o*PDA after 3 h of stirring at room temperature. For FTIR spectra, the assigned wavenumber range for each group is summarized from the reported polymer or oligomer forms of PDA.[23-26, 32, 36]

Cu (I) is very unstable in an aqueous environment with respect to disproportionation into Cu (0) and Cu (II). The presence of a large fraction of Cu (I) in the precursor solutions indicates that these Cu (I) are stabilized by the polymer or oligomer forms of PDA. The nitrogen atoms on C=N bond are found to coordinate with Cu (I) in all possible repeating units for polyPDA



(Figure S3), which agrees with the experimental observations.[24, 31, 37-38] Due to the complexity in the oxidative polymerization of PDA,[26] the detailed structure for Cu (I)-polyPDA complexes is difficult to characterize.

After 110 °C for 60 h, only the precursor solution containing $p$PDA produced Cu NWs (see the inset in Figure 4b). The successful synthesis of Cu NWs indicates that Cu (I)-polyPDA complexes are still redox-active in $p$PDA precursors. It has been shown that the charge transfer between transition metal ions and amines can be enhanced with increased temperature.[39] At 110 °C, the pre-reduced Cu (I) in the precursor solution is further reduced to Cu (0) through the coordinated nitrogen atoms. For these containing $m$PDA or $o$PDA, no Cu (0) diffraction was detected using XRD (Figure S4). It is consistent with the observed relatively weak reducing strength for $m$PDA or $o$PDA at room temperature (Figure 1 and 3a), at least in the precursor solution used in this study.

In order to verify whether the reducing strength is the reason for the differences in producing Cu NWs among the PDA isomers, an additional reducing agent, glucose, was added to the precursor solutions. As seen in Figure 4, with glucose, Cu NWs were obtained from the precursor solutions containing $p$PDA or $m$PDA within 24 h of reaction at 110 °C. Note that glucose is a mild reducing agent which does not reduce Cu (II) to Cu (0) at 110 °C.[15] Compared with Cu (II), Cu (I) is easier to be reduced to Cu (0) thermodynamically, with or without ligand.[40] A stabilized Cu (I) in the complex with polyPDA allows a mild reduction condition (using glucose) to produce Cu NWs. This is a great benefit to avoid using a harsh reduction environment, such as high temperature (> 165 °C), strong base and strong reducing agent (hydrazine etc.) as reported in literature.[14, 21, 39] The reducing strength of glucose, however, is compromised in the presence of $o$PDA due to a condensation reaction.[41] The synthesis results



from different reducing conditions are summarized in Table 1. Pre-reduction of Cu (II) to Cu (I)

is the key step to synthesize Cu NWs in a mild condition.

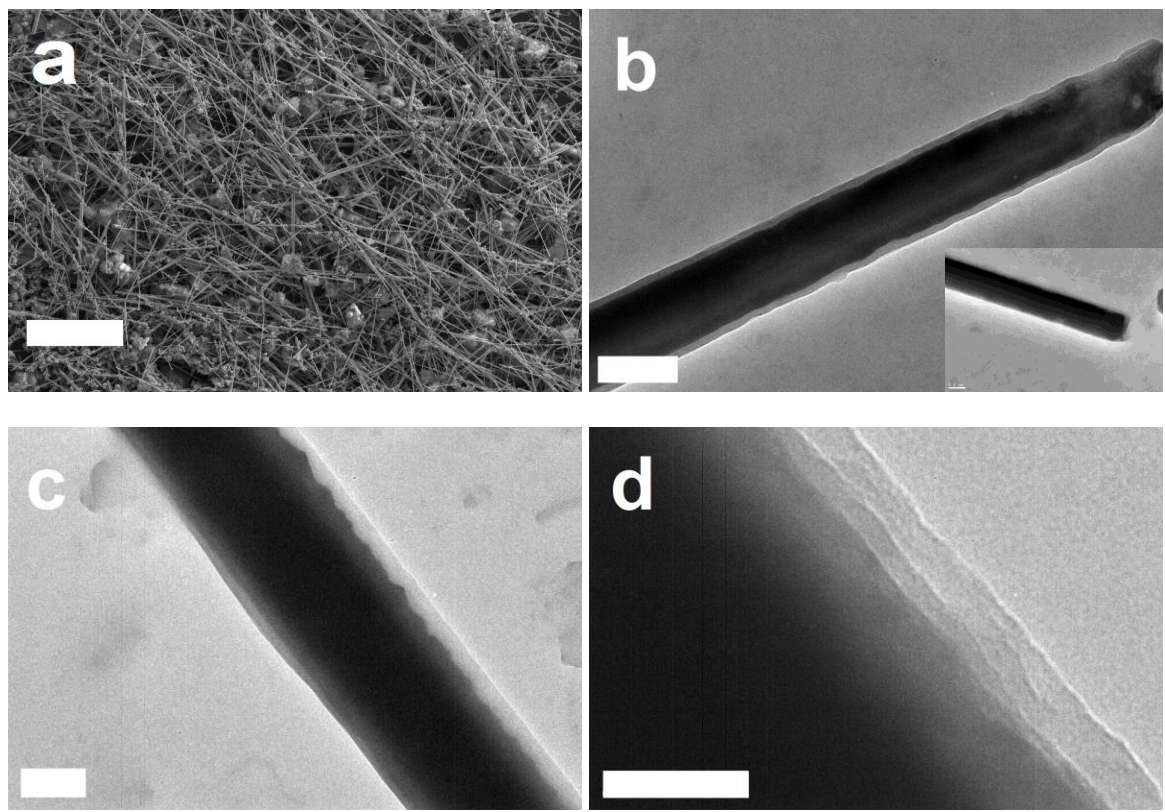

**Figure 4.** Microscopic images of the as-synthesized Cu NWs: (a) SEM of Cu NWs synthesized using *p*PDA (Scale bar = 20 µm); (b) TEM of Cu NWs synthesized using *p*PDA (Scale bar = 200 nm), and the inset for that synthesized without glucose; (c) TEM of Cu NWs synthesized using *m*PDA (Scale bar = 200 nm); (d) a polymer layer on the as-synthesized Cu NWs (Scale bar = 50 nm).



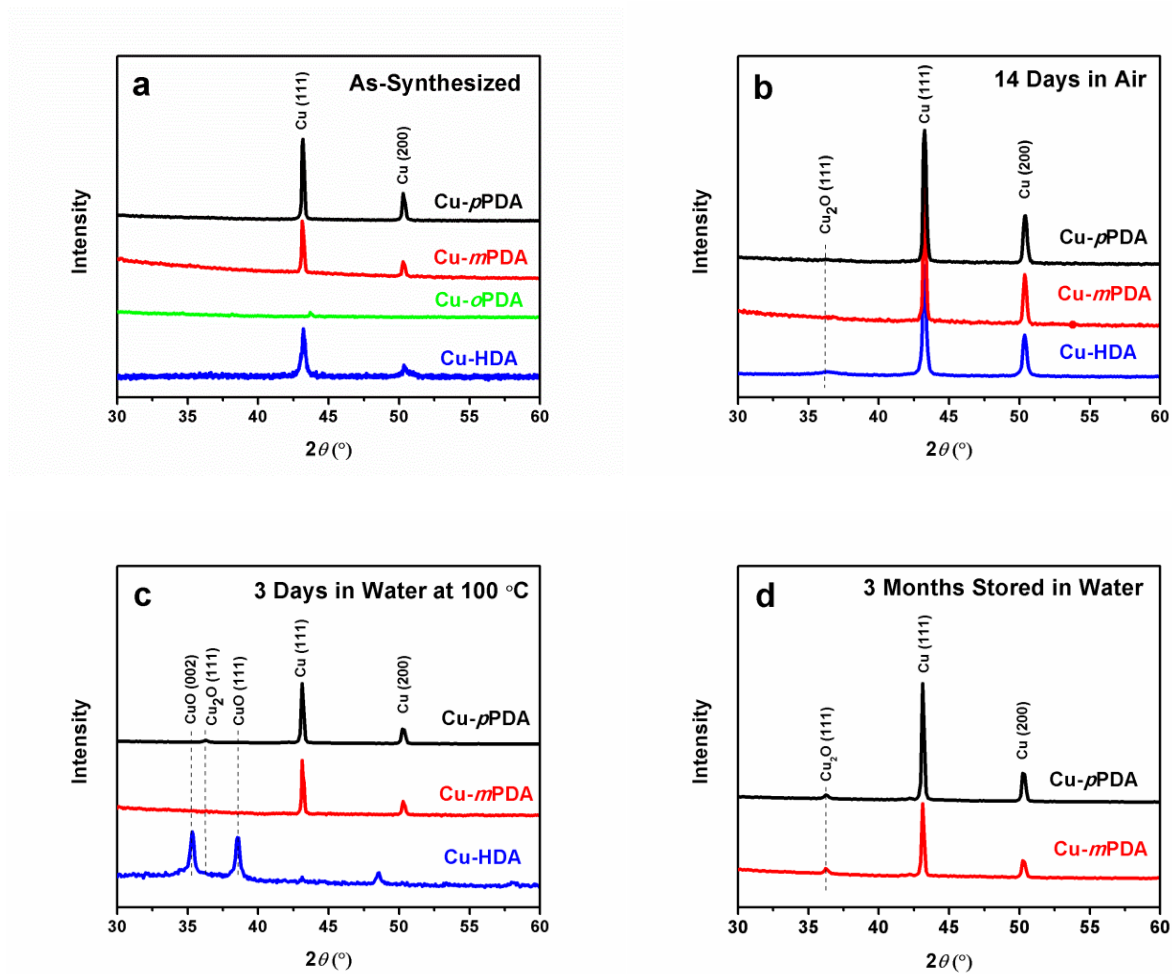

**Figure 5.** XRD patterns of Cu NWs synthesized using *p*PDA, *m*PDA, *o*PDA and HDA (a) synthesized without glucose; (b) synthesized with glucose; (c) after exposure to air for 14 days; (d) after treated in water at 100 °C for 3 days, and (e) after stored in water at room temperature for 3 months.

**Table 1.** Chemical reduction of Cu ions in an aqueous environment with different reducing conditions.

| PDA Isomer | 21 °C | 110 °C | 110 °C with glucose |
|---|---|---|---|
| *p*PDA | No Cu formation | Cu NWs | Cu NWs |
| *m*PDA | No Cu formation | No Cu formation | Cu NWs |
| *o*PDA | No Cu formation | No Cu formation | No Cu formation |





The XRD pattern, as seen in Figure 5a, confirms the presence of Cu with strong diffractions for Cu (111) and (200) facet. It is known that Cu NWs exhibit stronger diffractions for (111) facets while Cu nanoparticles have more (200) facets exposed.[42] The stronger (111) diffraction observed in the as-synthesized Cu NWs suggests a high yield of Cu NWs, which is similar to the high-quality Cu NWs synthesized using HDA.[3] The diameters for the as-synthesized Cu NWs with glucose are found to be $313 \pm 43$ and $475 \pm 60$ nm for $p$PDA and $m$PDA, respectively. Figure 4d show that these Cu NWs are coated with a thin layer of polyPDA.

The one-dimensional growth of Cu nanocrystals is likely due to the capping effect of PDA. Ethylenediamine (EDA) has been recently found to act as an etching agent instead of a capping agent, evidenced by very few EDA molecules were found on Cu NW surfaces.[19-20] If PDA followed the same mechanism as EDA, then the polymer shells should not form on Cu NW surfaces. It is possible that Cu atoms reduced in the polyPDA complexes diffused to the Cu seeds, and the coordinated polyPDA acts as a capping agent to ensure one dimensional growth. This process can proceed repeatedly to grow into Cu NWs.

The anti-oxidation properties of as-synthesized Cu NWs were compared with that synthesized using HDA. As seen in Figure 5b, no significant oxidation was observed for the Cu NWs after exposure to air at room temperature (21°C, humidity 40-60 %) for 14 days, except a small hump was observed for Cu NWs synthesized using HDA, which is the diffraction peak of $Cu_2O$ (111) facet. The high temperature and the presence of water is known to facilitate the oxidation process,[43] which limits the performance for Cu NWs in many applications, such as battery, catalysis and liquid sensors.  There are very few studies focusing on the stability in a hot environment with high humidity, because most of the Cu NWs can only be stable for a short





period of time in such a harsh condition. As we demonstrated in Figure 5c, HDA-synthesized Cu

NWs which were kept in boiling water (100°C) are completely oxidized to CuO within 3 days.

With a more densely packed dodecanethiol layer, Cu NWs can only be stable in an environment

with 85 % humidity at 85°C for 12 h. Under the same condition (3 days in boiling water), only a

small $Cu_2O$ peak was observed for the PDA-synthesized Cu NWs, suggesting an excellent

stability in harsh environment. Long-term stability was tested by storing the Cu NWs in water at

room temperature for 3 months. Figure 5d shows that only a small $Cu_2O$ (111) peak was

observed for both Cu NWs synthesized using *p*PDA and *m*PDA. For comparison, Cu NWs

synthesized from EDA were mostly oxidized after stored in water for only 1 month.[11] The

polyPDA coatings on Cu NW surfaces provide an efficient protection to prevent them from

oxidation in a high temperature aqueous environment.

The electrochemical properties of as-synthesized Cu NWs were tested using cyclic

voltammetry. PolyPDA does not have peaks in the scan range.[23] As shown in Figure 6, both Cu

NWs exhibit characteristic cathodic current peaks at -0.32 and -0.07 V, corresponding to Cu

(0)/Cu (I) and Cu (I)/Cu (II), respectively. The anode peaks for Cu (II)/Cu (I) and Cu (I)/Cu (0)

is at -0.50 and -0.81 V, respectively.[44] It suggests that the polyPDA coating does not inhibit the

electrochemical activity of Cu NWs. Unlike the aforementioned long-chain alkyl amines,

polyPDA unlikely protects Cu NWs through blocking the access for oxygen and water molecules.

PolyPDA allows the penetration of ions to some extent, demonstrated by their applications in

pseudo capacitor[23] and metal ion absorption,[24] therefore poly-PDA thin-film coated Cu NWs

remain electrochemical active, similar to that coated with polypyrrole.[21] The protection is

achieved likely through the electric field generated by the coordinated conducting polyPDA,

which restrict the electron flow[45]



The excellent oxidative stability makes these electrochemical active Cu NWs synthesized in this work a promising candidate for the applications in a high temperature, highly humid or aqueous environment, such as batteries and biosensors.

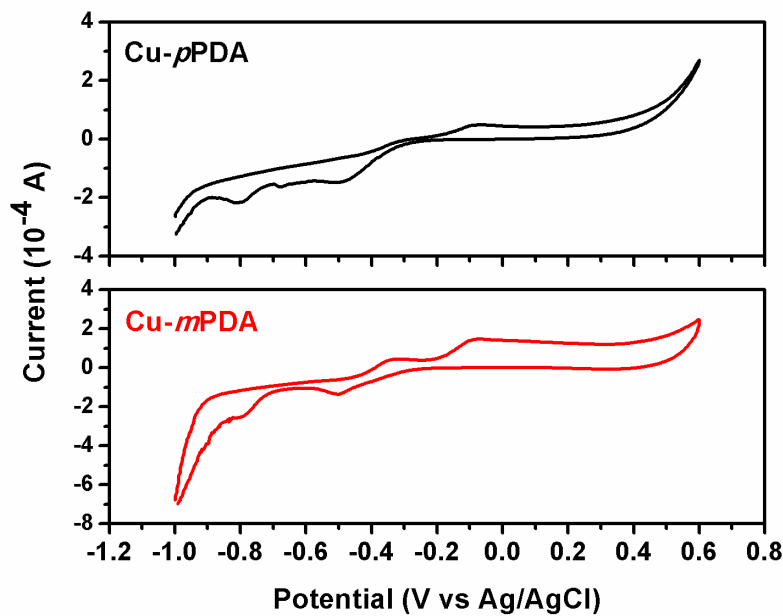

**Figure 6.** Cyclic voltammetry curves for Cu NWs synthesized using *p*PDA and *m*PDA in 0.1M KOH solution.

**Conclusion**

We have demonstrated the synthesis of Cu NWs using aromatic diamines for the first time. PDA has a multifunctional role in the Cu NW synthesis, which is coordinating, reducing and capping. PDA reduces Cu (II) to Cu (I) at room temperature, and stabilizes the resulting Cu (I) in an aqueous environment. The formation of stable Cu (I) complex is the key step to synthesize Cu NWs in a mild condition. Despite their identical chemical compositions, PDA isomers show different chemical reduction strength toward Cu (II). This indicates that not only the oxidation potential for PDA but also the coordination structure plays an important role in the chemical



reduction of Cu (II). The as-synthesized Cu NWs were covered with a thin layer of conducting

polyPDA polymer, which exhibited excellent anti-oxidation properties even in the presence of

water, suggesting the suitability of Cu NWs for applications in aqueous or highly humid

environment. The as-synthesized Cu NWs are electrochemical active, and can also be used for

biosensors, catalysis and supercapacitors because of the multifunctional nature of polyPDA. The

present work shows that using an aromatic diamine capping agent can provide effective surface

protection to the as-synthesized metal nanocrystals, which opens a new route for the core-shell of

metallic nanowires as a multi-functional building block for various nanotechnology applications.

**Acknowledgment**

The authors acknowledge the financial support from Lloyd's Registration Foundation, London,

UK, who funded this research through grants to protect life and property by supporting

engineering-related education, public engagement and the application of research. The authors

also thank Dr. Atif Aziz and Dr. Mark E. Welland from Nanoscience Centre at University of

Cambridge for the helpful discussions.